# CLASSICAL PHOTOABSORPTION SUM RULES*

STANLEY J. BRODSKY

*Stanford Linear Accelerator Center*
*Stanford University, Stanford, California 94309*

and

IVAN SCHMIDT

*Universidad Federico Santa María*
*Casilla 110–V, Valparaíso, Chile*

## ABSTRACT

We use a quantum loop expansion to derive sum rule constraints on polarized photoabsorption cross sections in the Standard Model, generalizing earlier results obtained by Altarelli, Cabibbo, and Maiani. We show that the logarithmic integral of the spin-dependent photoabsorption cross section $\int_{\nu_{th}}^{\infty} \frac{d\nu}{\nu} \Delta\sigma_{\text{Born}}(\nu)$ vanishes for any $2 \to 2$ process $\gamma a \to bc$ in the classical, tree-graph approximation in the Standard Model, supersymmetric gauge theories, and other quantum field theories where the Drell-Hearn Gerasimov sum rule is valid. Here $\nu = p \cdot q/M$ and $\Delta\sigma(\nu) = \sigma_P(\nu) - \sigma_A(\nu)$ is the difference between the photoabsorption cross section for parallel and antiparallel photon and target helicities. Tests of the sum rule for the reactions $\gamma e \to W\nu$ and $\gamma\gamma \to W^+W^-$ can provide new tests of the canonical magnetic and quadrupole couplings of the Standard Model. We also extend the sum rule to certain virtual photon processes.

Submitted to *Physics Letters B.*

★ Work supported in part by Department of Energy contracts DE–AC03–76SF00515 and DE–AC02–76ER03069, and by Fondo Nacional de Investigación Científica y Tecnológica, Chile, contract 1931120.

## I. Introduction

The Dirac value $g = 2$ for the magnetic moment $\mu = geS/2M$ of a particle of charge $e$, mass $M$, and spin $S$, plays a special role in quantum field theory. As shown by Weinberg [1] and Ferrara *et al.* [2], the canonical value $g = 2$ gives an effective Lagrangian which has maximally convergent high energy behavior for fields of any spin. In addition, one can use the Drell-Hearn Gerasimov (DHG) sum rule [3] to show that the magnetic and quadrupole moments of spin-$\frac{1}{2}$ or spin-1 bound states approach the canonical values $\mu = eS/M$ and $Q = -e/M^2$ in the zero radius limit $MR \to 0$ [4, 5, 6], independent of the internal dynamics.

In the case of the Standard Model, the anomalous magnetic moments $\mu_a = (g-2)eS/2M$ and anomalous quadrupole moments $Q_a = Q + e/M^2$ of the fundamental fields vanish at tree level, ensuring a quantum field theory which is perturbatively renormalizable. The canonical values $\mu = eS/M$ and $Q = -e/M^2$ are thus fundamental properties of the classical limit of the Standard Model. The anomalous couplings receive contributions of order $\alpha/\pi$ from radiative corrections [7]. Deviations from the predicted values could reflect new physics and interactions such as virtual corrections from supersymmetry or an underlying composite structure.

The canonical values $g = 2$ and $Q = -e/M^2$ lead to a number of important phenomenological consequences:

1. The magnetic moment of a particle with $g = 2$ processes with the same frequency as the Larmor frequency in a constant magnetic field. This synchronicity is a consequence of the fact that the electromagnetic spin currents can be formally generated by an infinitesimal Lorentz transformation [8, 9].

2. The forward helicity-flip Compton amplitude for a target with $g = 2$ vanishes

at zero energy [10].

3. The amplitude for a photon radiated in the scattering of any number of incoming and outgoing particles with charge $e_i$ and four-momentum $p_i^\mu$ vanishes at the kinematic angle where all the ratios $e_i/p_i \cdot k$ are simultaneously equal [9]. For example, the Born cross section $d\sigma/\cos\theta_{cm}(u\overline{d} \to W^+\gamma)$ vanishes identically at an angle determined from the ratio of charges: $\cos\theta_{cm} = e_d/e_{W^+} = -1/3$ [11]. Such "radiative amplitude zeroes" or "null zones" occur at lowest order in the Standard Model because the electromagnetic spin currents of the quarks and the vector gauge bosons are all canonical.

The vanishing of the forward helicity-flip Compton amplitude at zero energy for the canonical couplings, together with the optical theorem and dispersion theory, leads to a superconvergent sum rule; *i.e.*, a zero value for the DHG sum rule. This remarkable observation was first made for quantum electrodynamics and the electroweak theory by Altarelli, Cabibbo and Maiani [12]. In this paper we shall use a quantum loop expansion to show that the logarithmic integral of the spin-dependent part of the photoabsorption cross section

$$\int_{\nu_{th}}^{\infty} \frac{d\nu}{\nu} \Delta\sigma_{\text{Born}}(\nu) = 0 \qquad (1)$$

for any $2 \to 2$ Standard Model process $\gamma a \to bc$ in the classical, tree graph approximation. The particles $a, b, c$ and $d$ can be leptons, photons, gluons, quarks, elementary Higgs particles, supersymmetric particles, etc. We also can extend the sum rule to certain virtual photon processes. Here $\nu = p \cdot q/M$ is the laboratory energy and $\Delta\sigma(\nu) = \sigma_P(\nu) - \sigma_A(\nu)$ is the difference between the photoabsorption cross section for parallel and antiparallel photon and target helicities. The sum rule receives nonzero contributions in higher order perturbation theory in the Standard Model from both quantum loop corrections and higher particle number final states.

We shall refer to the superconvergent relation Eq. (1) as a classical photoabsorption sum rule. More generally, we can use Eq. (1) as a new way to test the canonical couplings of the Standard Model and to isolate the higher order radiative corrections. The sum rule also provides a non-trivial consistency check on calculations of the polarized cross sections. Probably the most interesting application and test of the Standard Model is to the reactions $\gamma e \to W \nu$ and $\gamma e \to Z e$ which can be studied in high energy polarized electron-positron colliders with back-scattered laser beams. In contrast to the timelike process $e^+ e^- \to W^+ W^-$, the $\gamma e$ reactions are sensitive to the anomalous moments of the gauge bosons at $q^2 = 0$. The Standard Model predicts

$$\int_{\nu_{th}}^{\infty} \frac{d\nu}{\nu} \Delta \sigma_{\gamma e \to W \nu}(\nu) = \mathcal{O}(\alpha^3). \tag{2}$$

The vanishing of the logarithmic integral of $\Delta \sigma(\nu)$ at the tree-graph approximation also implies that there must be an energy $\nu_0$ where $\Delta \sigma_{\text{Born}}(\nu_0) = 0$ [13]. Modifications of the Standard Model, such as those arising from composite structure of the quarks or vector bosons, will lead to corrections to the sum rule. We will investigate the sensitivity of the position of the crossing point $\sqrt{s}_{\gamma e} \simeq 3.16 M_W$ and the value of the DHG integral to higher order corrections and violations of the Standard Model in Ref. [13].

## II. Classical Limit of the DHG Sum Rule

The DHG sum rule relates the square of the anomalous magnetic moment of a particle to the logarithmic integral of the difference of the cross section for the absorption of a photon with spin parallel or antiparallel to the target spin:

$$\mu_a^2 = \frac{4\pi\alpha S^2}{M^2}(g-2)^2 = \frac{S}{\pi}\int_{\nu_{th}}^{\infty}\frac{d\nu}{\nu}\left[\sigma_P(\nu) - \sigma_A(\nu)\right]. \qquad (3)$$

The DHG sum rule is derived by assuming an unsubtracted dispersion relation for the forward spin-flip Compton amplitude $f_2(\nu)$, plus the low energy theorem which relates this amplitude at zero energy to the square of the anomalous moment [10]. It is valid for a target of any spin $S$, whether elementary or composite [3, 6].

Some years ago Weinberg noticed that the validity of the DHG sum rule requires that the first term in a perturbative expansion of the anomalous magnetic moment is of order $\alpha$, so that $g = 2$ at tree level for a particle of any spin [1]. This can be easily seen by expanding both sides of Eq. (3) in powers of $\alpha$. The lefthand side of the DHG sum rule would start at order $\alpha$ if the perturbative expansion for the anomalous magnetic moment had a contribution of order $\alpha^0$. But the first contribution to the righthand side is elastic Compton scattering, which starts at order $\alpha^2$. Consistency then demands Weinberg's result. Conversely, the Standard Model predicts that the lefthand side starts as $\alpha^3$. Thus there is another important consequence of the DHG sum rule: the tree graph contributions of order $\alpha^2$ to the DHG integral must vanish. This can be easily verified in QED [12] by explicit integration of the logarithmic integral of the Compton scattering cross section $\gamma e \to \gamma e$ in Born approximation:

$$\sigma_P(\nu) - \sigma_A(\nu) = -\frac{2\pi\alpha^2}{m\nu}\left[\left(1+\frac{m}{\nu}\right)\ell n\left(1+\frac{2\nu}{m}\right) - 2\left(1+\frac{\nu^2}{(m+2\nu)^2}\right)\right]. \qquad (4)$$

We can extend these results to general processes in quantum field theory by considering a loop expansion of the perturbative contributions to the forward Compton amplitude and demanding that both sides of the dispersion relation track in the number of loops. The classical limit corresponds to zero quantum loops. Loops can be counted by the following device [14]: to each vertex of a Feynman graph one assigns a factor of $a^{-1}$. To each propagator one assigns a factor of $a$, since it is the inverse of the differential operator that appears in the kinetic term in the Lagrangian $\mathcal{L}$. If $P$ is the number of propagators and $V$ is the number of vertices, then $P - V$ counts the powers of $a$ and $L = P - V + 1$ is the number of loops of the Feynman graph. This counting corresponds to formally rescaling $\mathcal{L} \to \mathcal{L}\, a^{-1}$. The rescaling changes the quantum theory but not the classical theory.

In the rescaled quantum theory, the dispersion relation for the forward spin-flip Compton amplitude $f_2(\nu)$ is modified by a power of $a$:

$$a\mu_a^2 \propto \int_{\nu_{th}}^{\infty} \frac{d\nu}{\nu} \mathrm{Im} f_2(\nu). \qquad (5)$$

The extra power of $a$ appears on the left hand side since the zero energy spin-flip Compton amplitude $f_2(0)$ contains one extra propagator not counted in the square of the anomalous moment [15].

The Schwinger $\alpha/2\pi$ contribution to the anomalous magnetic moment of the electron is computed in QED from a Feynman amplitude with $V = 3, P = 3, L = 1$. Thus the lefthand side of the dispersion relation Eq. (5) is proportional to $a^1$. On the other hand, the lowest order (box graph) contribution to $\mathrm{Im} f_2(\gamma e \to \gamma e)$ has $V = 4, P = 4, P - V = 0$, i.e., the power $a^0$, so it must have a vanishing integral. Thus, by the optical theorem, the Born contribution to $\Delta \sigma(\gamma e \to \gamma e)$ cannot contribute to the DHG sum rule. Notice that the two-loop contribution to $\mathrm{Im} f_2(\nu)$ has $V = 6, P = 7, P - V = 1$; it thus has power $a^1$ and can contribute to the DHG integral. The two-loop contribution to $\mathrm{Im} f_2(\nu)$ has a cut which corresponds to the square of the tree graph contribution to $\sigma(\gamma e \to \gamma\gamma e)$ as well as virtual radiative corrections to $\sigma(\gamma e \to \gamma e)$. Thus $2 \to 3$ tree graphs can give non-zero contributions to the DHG integral, but the contributions of $2 \to 2$ tree graphs to the logarithmic integral of $\Delta\sigma(\gamma e \to \gamma e)$ vanish.

The above counting of quantum loops reflects the non-linearity of the forward Compton dispersion relation and is representative of all Standard Model contributions to the DHG integral. In fact, in any field theory where the DHG sum rule is valid, the tree graph contributions to $2 \to 2$ processes give zero DHG integral; *i.e.*

$$\int_{\nu_{th}}^{\infty} \frac{d\nu}{\nu} \left[ \sigma_P^{\gamma a \to bc}(\nu) - \sigma_A^{\gamma a \to bc}(\nu) \right]_{\mathrm{Born}} = 0. \tag{6}$$

### III. Sum Rule Tests of the Standard Model

We now consider some explicit examples of the classical photoabsorption sum rule Eq. (6) in the Standard Model.

1. $\gamma e \to W\nu$. The cross section has been computed in the Standard Model at lowest order for polarized photons and massless electrons in Ref. [16]. The result for the difference of cross sections for parallel and antiparallel helicities is:

$$\Delta\sigma = \frac{2\widetilde{\sigma}}{x} \left[ \frac{x-1}{4x} \left( 13 + \frac{3}{x} \right) - \left( 1 + \frac{3}{x} \right) \ell n\, x \right], \tag{7}$$

where $\widetilde{\sigma} = \pi\alpha^2/M_W^2 \sin^2\theta_W \simeq 47$ pb, and $x = s_{\gamma e}/M_W^2$. It is convenient to change variables to $y = 1/x$. The DHG integral is then proportional to:

$$\int_0^1 dy \left[ (1-y)(13+3y)/4 + (1+3y)\ell n y \right], \tag{8}$$

which indeed does vanish. This result was first verified in Ref. [12]. The cancellation of the positive and negative contributions of $\Delta\sigma(\gamma e \to W\nu)$ to the DHG integral is evident in Fig. 1.

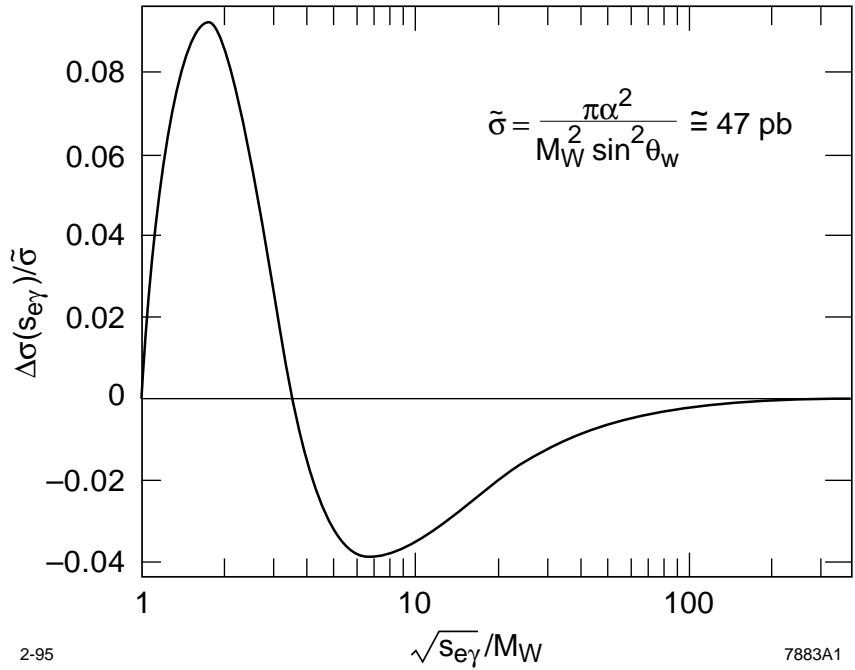

Figure 1. The Born cross section difference $\Delta\sigma$ for the Standard Model process $\gamma e \to W\nu$ for parallel minus antiparallel electron/photon helicities as a function of $\log\sqrt{s}_{e\gamma}/M_W$. The logarithmic integral of $\Delta\sigma$ vanishes in the classical limit.

2. The classical photoabsorption sum rule also applies to photon targets, and thus to $\gamma\gamma$ annihilation to any two-body final state such as $\gamma\gamma \to \ell\bar{\ell}$, $\gamma\gamma \to ZZ$, $\gamma\gamma \to HH$ and $\gamma\gamma \to q\bar{q}$ at tree level. An interesting example is the process $\gamma\gamma \to W^+W^-$. The Born approximation cross section has been computed by Ginzburg et al. [16]. The difference in cross sections for the annihilation of photons with total helicity 0 and 2 is

$$\Delta\sigma = \frac{\tilde{\sigma}v}{8x}\left[-19 + \left(8 - \frac{5}{x}\right)L\right], \qquad (9)$$

where $\tilde{\sigma} = 8\pi\alpha^2/M_W^2 \simeq 86$ pb, $x = s_{\gamma\gamma}/(4M_W^2)$, $v = \sqrt{1-1/x}$, and $L = \frac{1}{v}\ell n\frac{1+v}{1-v}$. Changing variables to $y = 1/x$, we obtain a DHG integral proportional to:

$$\int_0^1 dy \left[-19v + (8-5y)\ell n \frac{1+v}{1-v}\right] = 0, \qquad (10)$$

where $v = \sqrt{1-y}$. As expected, the DHG integral vanishes for the tree-graph cross section.

Classical photoabsorption sum rules can also be derived for processes involving off-shell photons. Consider the contribution of the QED process $\Delta\sigma(\gamma e \to \mu^+\mu^- e)$ to the DHG sum rule for the electron anomalous magnetic moment. Muon loops contribute to the lefthand side of the sum rule at order $a^2$ and higher via vacuum polarization and light-by-light scattering contributions to the electron anomalous moment interfering with the one-loop Schwinger contribution. On the other hand, the contribution to the forward spin-flip Compton amplitude $\text{Im} f_2(\nu)$ from the cut contribution $\gamma e \to \mu^+\mu^- e$ first appears on the righthand side at order $a$; thus the tree-graph contribution to the DHG integral for this process must also vanish. The same argument also implies that the DHG integral vanishes for virtual photoabsorption processes such as $\ell\gamma \to \ell Q\overline{Q}$ and $\ell g \to \ell Q\overline{Q}$, the lowest order sea-quark contribution to polarized deep inelastic photon and hadron structure functions. Note that in each case the integral extends to $\nu = \nu_{th}$, which is generally beyond the usual leading twist domain. These results can clearly be generalized to other higher order tree-graph processes in the Standard Model and supersymmetric gauge theory.

ACKNOWLEDGMENTS

We would like to thank Sid Drell, Olivier Espinosa, Fred Gilman, Michael Peskin, Tom Rizzo, and David Whittum for helpful conversations.